\documentclass[12pt,preprint]{aastex}   
\def\paren#1{\left( #1 \right)}

\def\Mesz{M\'esz\'aros~}
\begin{document}
\title{Inverse Compton X-ray Flare from GRB Reverse Shock}
\author{S. Kobayashi\altaffilmark{1,2,3}, 
        B. Zhang \altaffilmark{4},
        P. \Mesz \altaffilmark{1,2} and
        D. Burrows \altaffilmark{2}}
\altaffiltext{1}{Center for Gravitational Wave Physics,
                 Pennsylvania State University, University
		 Park, PA 16802}
\altaffiltext{2}{Department of Astronomy and Astrophysics, 
                 Pennsylvania State University, University
		 Park, PA 16802}
\altaffiltext{3}{Astrophysics Research Institute, 
                 Liverpool John Moores University, 
		 Birkenhead CH41 1LD UK}
\altaffiltext{4}{Department of Physics,
                 University of Nevada, Las Vegas,
		 NV 89154}
\begin{abstract}
 We study synchrotron self-inverse Compton radiation from a
 reverse-shocked fireball. If the inverse Compton process dominates the 
 cooling of shocked electrons, an X-ray flare produced by the first
 order Compton scattering would emerge in the very early afterglow phase, 
 with the bulk of the shock energy radiated in the second order 
 scattering component at 10-100 MeV. The dominance of inverse Compton 
 cooling leads to the lack of prompt optical flashes. 
 We show that for plausible parameters this scattering process
 can produce an X-ray flare with a relative amplitude change by a factor
 of several. Flares with a larger amplitude and multiple X-ray flares in
 a single event are likely to be produced by another mechanism
 (e.g. internal shocks). 
\end{abstract}
\section{Introduction}
The Swift satellite is a multi-wavelength
observatory designed to detect GRBs and their X-ray and UV/optical
afterglows. Thanks to its fast pointing capabilities Swift is disclosing
the early afterglow phase. The Swift X-Ray Telescope (XRT) found that
most X-ray afterglows fall off rapidly for the first few hundred
seconds, followed by a less rapid decline (Tagliaferri et al. 2005). In
the early afterglows of GRB 050406 and GRB 050502b, XRT detected
mysterious strong X-ray flares: rapid brightening of the X-ray afterglow
after a few hundred seconds post-burst (Burrows et al. 2005). These
results suggest the existence of additional emission components in the early
afterglow phase besides the conventional forward shock emission. 

In recent years, ground-based robotic telescopes reported the lack of
prompt optical emission, except for a few cases. The Swift UV/Optical
Telescope (UVOT) provided further stringent upper limits at very early
epochs after the bursts (Roming et al. 2006). The limits $\sim 20$ mag at
$\lesssim 100$ sec are much lower than the 9th magnitude 
optical flash associated with GRB 990123 (Akerlof et al. 1999). Although
the host extinction at $z\sim 1$ and/or the high magnetization of a fireball
(Zhang \& Kobayashi 2005) can explain the lack of optical flashes,
other suppression mechanisms might be involved. 

According to the standard relativistic fireball model, reverse shocks are
expected to radiate photons in the optical/IR band via the synchrotron process
in a very early afterglow phase (\Mesz \& Rees 1997; Sari \& Piran
1999). Although the contribution of the reverse 
shock synchrotron emission to the X-ray band is small, electrons in the
reverse-shocked region can up-scatter the synchrotron photons 
(synchrotron self-inverse Compton emission, hereafter SSC)
to the X-ray or even higher energies 
\footnote{
Previous studies mainly address the prompt high-energy gamma-rays 
($>$ 100 MeV) detected by EGRET. Higher order scattering was ignored.
As we will show, for plausible parameters the second order IC is
important to discuss early afterglow. The correction significantly 
modifies the amplitude of an X-ray flare (the first order IC component) when 
the Compton parameter is large.} (\Mesz \& Rees 1993; Wang, Dai \&
Lu 2001a,b; Granot \& Guetta 2003).
The SSC emission should produce
additional features in the early X-ray afterglows. If inverse-Compton 
(IC) cooling is the dominant cooling mechanism of the electrons, 
the energy available for the synchrotron process is significantly 
reduced, and prompt optical flashes could be faint
(Beloborodov 2005). In this paper, we study the role of the 
reverse shock SSC emission in early X-ray afterglows.
\section{Reverse Shocked Ejecta}
We consider a relativistic shell (fireball ejecta) with an isotropic 
energy $E$ and an initial Lorentz factor $\Gamma_{0}$ expanding into a
homogeneous ambient medium with particle number density $n$.  
The evolution of reverse shocks is
classified into two cases (Sari \& Piran 1995) by using a critical
Lorentz factor
$\Gamma_c=\paren{3(1+z)^3E/32\pi nm_pc^5 T^3}^{1/8} 
\sim 130 ~\zeta^{3/8} E_{52}^{1/8} T_2^{-3/8} n^{-1/8}$
where $\zeta=(1+z)/2$, $E_{52}=E/10^{52}$ ergs, $T_2=T/10^2$ sec and 
$T$ is the duration of the GRB.
If $\Gamma _0 >\Gamma_c$ (thick shell case), the shell is significantly
decelerated by a reverse shock. The Lorentz factor at the shock crossing
time $t_d \sim T$ is given by $\Gamma_d \sim \Gamma_c$. 
If $\Gamma_0 < \Gamma_c$ (thin shell case), the reverse
shock can not decelerate the shell effectively and $\Gamma_d \sim
\Gamma_0$. The deceleration time $t_d=(1+z)(3E/32\pi \Gamma_0^8n m_p
c^5)^{1/3}$ is larger than the GRB duration. 

We assume that constant fractions $\epsilon_e$ and $\epsilon_B$ of the
shock energy go into the electrons and magnetic fields, respectively, 
and that electrons are accelerated in the
shock to a power-law distribution $Nd\gamma \propto \gamma^{-p}d\gamma,
\gamma\ge\gamma_m \sim 
[(p-2)/(p-1)]\epsilon_e (m_p/m_e) (\Gamma_0/\Gamma_d)
\sim 180 (\epsilon_e/0.3)(\Gamma_0/\Gamma_d)$
where $p=2.5$ was assumed. Using the number of electrons in
the shell $N_e$ and the deceleration radius $R_d \sim 2c\Gamma_d^2 t_d/(1+z)$,
the optical depth of the shell is 
$\tau(R_d) =\sigma_T N_e/4\pi R_d^2
          =(1/3)\mathcal{R}_M\sigma_T n R_d$
where we used the fact that the mass of the shell is larger by 
a factor of $\mathcal{R}_M=\Gamma_d^2/\Gamma_0$ than that of 
the ambient material swept by the shell at the deceleration time.

\section{The First and Second order IC}
Reverse-shocked electrons emit optical photons via the synchrotron 
process. Since the random Lorentz factors of electrons are 
typically a few hundreds, the up-scattered
photons are in X-ray band. If the first order IC is in the X-ray band 
($h\nu_X \sim$ 5 keV), the comoving photon energy of the first IC is
$h\nu_X^\prime=h\nu_X/\Gamma \sim 50$ eV where $\Gamma$ is the bulk
Lorentz factor of the shocked shell. The Klein-Nishina effect does
not suppress the second order IC scattering as long as the random
Lorentz factor $\gamma_m$ is below $\sim m_ec^2/h\nu_X^\prime \sim 100\Gamma$.
The second IC component appears in $10-100$ MeV range. 
Higher order scattering (three or more) can be ignored because of the 
Klein-Nishina effect.

The ratio of the IC to synchrotron luminosity can be computed in a 
general way (Sari \& Esin 2001). The luminosity ratios, in the limit of
up to second scattering, are given by 
\begin{eqnarray}
x&\equiv& \frac{L_{IC,1st}}{L_{syn}}=\frac{U_{syn}}{U_B}
=\eta \frac{U_e}{U_B}\paren{1+\frac{U_{IC,1st}}{U_{syn}}
+\frac{U_{IC,2nd}}{U_{syn}}}^{-1}
=\eta\frac{\epsilon_e}{\epsilon_B}\paren{1+x+x^2}^{-1}
\label{eq:x1}
\\
x_2&\equiv& \frac{L_{IC,2nd}}{L_{syn}}=\frac{U_{IC,1st}}{U_B} 
=\frac{U_{IC,1st}}{U_{syn}} \frac{U_{syn}}{U_B}=x^2 
\label{eq:x2}
\end{eqnarray}
where $U_{syn},U_{IC,1st},U_{IC,2nd}, U_B$ and $U_e$ are the 
energy density of synchrotron radiation, the 1st IC, 2nd IC,
magnetic field and random electrons, respectively. 
We used $\eta U_e=U_{syn}+U_{IC,1st}+U_{IC,2nd}$. $\eta$ is the 
fraction of the electron energy that was radiated away: 
$\eta=1$ for fast cooling and $\eta=(\gamma_c/\gamma_m)^{2-p}$
for slow cooling. Solving eq (\ref{eq:x1}) for $x$ we obtain
\begin{equation}
x=
\left\{
\begin{array}{@{\,}ll}
 (\eta \epsilon_e/\epsilon_B)  &  
    \mbox{if ~}   (\eta \epsilon_e/\epsilon_B) \ll 1 \\    
 (\eta \epsilon_e/\epsilon_B)^{1/3}  &   
 \mbox{if ~} (\eta \epsilon_e/\epsilon_B) \gg 1  
\end{array}
\right. 
\label{eq:solx1}
\end{equation} 
When the IC emission dominates the overall electron cooling,
it reduces the energy available for synchrotron radiation. Consequently,
the cooling-break energy of the electron distribution $\gamma_c$ is reduced
from the synchrotron-only value $\gamma_{c,s}$ by a factor of
$(1+x+x^2)$.

If $(\epsilon_e/\epsilon_B) \gg 1$ and  
$\eta_{s}\equiv (\gamma_{c,s}/\gamma_m)^{2-p} \ll 1$, the efficiency
$\eta$ depends on $x$. From eq. (\ref{eq:x1}), one obtains
$x\sim  (\eta_s\epsilon_e/\epsilon_B)$ for 
$(\epsilon_e/\epsilon_B) < r_1$, 
$x\sim (\eta_s \epsilon_e/\epsilon_B)^{1/(7-2p)}$ for
$r_1 < (\epsilon_e/\epsilon_B) < r_2$, and 
$x\sim (\epsilon_e/\epsilon_B)^{1/3}$ for 
$(\epsilon_e/\epsilon_B) > r_2$
where $r_1=\eta_s^{-1}$ and $r_2=\eta_s^{-3/(2p-4)}$.
A smaller $\epsilon_B$ makes the Compton parameter $x$ larger, but the
dependence is rather weak. Considering scalings 
$\eta_s \propto \epsilon_B^{p-2}$,
$r_1 \propto \epsilon_B^{2-p} \sim \epsilon_B^{-1/2}$ and
$r_2 \propto \epsilon_B^{-3/2}$, the second regime
$r_1 < (\epsilon_e/\epsilon_B) < r_2$ should be achieved
in the limit of $\epsilon_B\to 0$. Then, we obtain 
$x \propto \epsilon_B^{-(3-p)/(7-2p)} \sim \epsilon_B^{-1/4}$ 
for $p=2.5$.

\section{SSC Emission from Reverse Shock}

The synchrotron spectrum of reverse shock emission can be approximated 
by a broken power-law with break frequencies,  typical frequency
$\nu_{m}$  and cooling frequency $\nu_{c}$ (e.g. Sari, Piran \& Narayan
1998). The 1st IC spectrum is also roughly described by a broken power
law with break frequencies $\nu_{m}^{IC}$ and $\nu_{c}^{IC}$. The spectral
characteristics of the 1st SSC emission are given by 
\begin{equation}
\nu_{m}^{IC} \sim   2\gamma_{m}^2 \nu_{m}, \ \
\nu_{c}^{IC} \sim   2\gamma_{c}^2 \nu_{c}, \ \
F_{max}^{IC} \sim   \kappa \tau F_{max},
\label{eq:IC}\\
\end{equation}
where $\kappa$ is a correction coefficient. 
The ratio of $\nu F_\nu$ peaks also gives their luminosity ratio. It is
$L_{IC,1st}/L_{syn} 
\sim \nu F_\nu^{IC}(\nu_{peak}^{IC})/
\nu F_\nu(\nu_{peak})
= 2\kappa \tau \gamma_m\gamma_c \eta$
where $\nu_{peak}=\max[\nu_m,\nu_c]$ and 
$\nu^{IC}_{peak}=\max[\nu_m^{IC},\nu_c^{IC}]$.
We can reduce this estimate to eq. (\ref{eq:solx1})
with a normalization $\kappa = 4(p-1)/(p-2)$.

To assess the relative importance of the 1st IC emission and the
synchrotron emission in the X-ray band (the 2nd IC component is well
above X-ray band), we consider the flux ratio at the IC ($F_\nu$) peak.
For slow cooling, the IC spectrum peaks at $\nu_m^{IC}$, and declines as   
$F_\nu^{IC} \propto \nu^{-(p-1)/2}$ for $\nu_m^{IC} <\nu<\nu_c^{IC}$ 
and as $F_\nu^{IC} \propto \nu^{-p/2}$ above $\nu_c^{IC}$, while
the synchrotron spectrum peaks at $\nu_m$, and declines as  
$F_\nu \propto \nu^{-(p-1)/2}$ for $\nu_m<\nu<\nu_c$ and as 
$F_\nu \propto \nu^{-p/2}$ above $\nu_c$. The flux ratio at the 
$F_\nu^{IC}$ peak is given by  
\begin{equation}
\frac{F_\nu^{IC}(\nu_m^{IC})}{F_\nu(\nu_m^{IC})} 
=\frac{F^{IC}_{max}}{F_\nu(\nu_m^{IC})} 
\lesssim \gamma_m^{p-2}x
\label{eq:fluxratio}
\end{equation}
where $p<3$ was assumed. The effective Compton parameter $x= 2
\kappa \tau \gamma_m\gamma_c \eta$; the equality is achieved only when
$\nu_c \sim \nu_m$. If electrons are in the fast cooling regime, the 1st IC 
peaks at $\nu_c^{IC}$. We can show that the flux ratio at 
$\nu_c^{IC}$ has the same upper limit  $\lesssim \gamma_m^{p-2}x$.
Therefore, the contrast of a SSC bump should be 
less than $\gamma_m^{p-2}x \sim 10x ~(\epsilon_e/0.3)^{p-2} (\Gamma_0/\Gamma_d)^{p-2}$  
where $(\Gamma_0/\Gamma_d) \sim 1$ for a thin shell. The bump could 
be more significant in the thick shell case, but in this case 
the SSC emission occurs
at the end of the prompt gamma-ray emission, and it is difficult to 
separate the SSC emission from the internal shock signal. 

If the $\epsilon_B$ parameter is small, the IC cooling becomes more 
important compared to the synchrotron cooling. However, as we showed at
the end of the previous section, the dependence is weak: $x\propto
\epsilon_B^{-1/4}$. To achieve $x \gg 1$, a very small $\epsilon_B$ is
required. The cooling frequency $\nu_c$ then becomes much higher than the
typical frequency $\nu_m$. It is possible to show that the flux ratio 
at $\nu_m^{IC}$ is insensitive to $\epsilon_B$ as
$F_\nu^{IC}(\nu_m^{IC})/F_\nu(\nu_m^{IC})=
(F_{max}^{IC}/F_{max})(\nu_m^{IC}/\nu_m)^{(p-1)/2}
\sim  x\gamma_c^{p-3} \propto \epsilon_B^0$.
For plausible parameters, $x$ could be several at most.
If $x$ is large, most energy is radiated in the 2nd IC component, the SSC
bump and the base line X-ray afterglow should be faint. The synchrotron
emission (optical flash) should be highly suppressed.

\section{Forward Shock Emission}
A forward shock also emits X-rays via the synchrotron process.
At the deceleration time, the
spectral characteristics of the forward and reverse shock synchrotron
emission are related (e.g. Zhang, Kobayashi \& \Mesz 2003) as 
\begin{equation}
\nu_{m,f} \sim \mathcal{R}_B^{1/2} \mathcal{R}_M^2    \nu_{m}, \ \  
\nu_{c,f} \sim \mathcal{R}_B^{-3/2}\mathcal{R}_X^{-2} \nu_c, \ \
F_{max,f} \sim \mathcal{R}_B^{1/2} \mathcal{R}_M^{-1}  F_{max}
\label{eq:zkm03}
\end{equation}
where the subscript $f$ denotes forward shock quantities,
we have assumed that the $\epsilon_e$ and $p$ parameters are the same
for both shock regions, but with different $\epsilon_B$ as parameterized
by $\mathcal{R_B}=(\epsilon_{B,f}/\epsilon_B)$. The reason 
we introduce the $\mathcal{R_B}$ parameter is that the fireball ejecta
could be endowed with a primordial magnetic field (e.g. Zhang et al. 2003 and 
references therein). $\mathcal{R}_X=(1+x_f)/(1+x+x^2)$ is a
correction factor for the IC cooling. In the forward shock region, a
once-scattered synchrotron photon generally 
has energy larger than the electron mass in the rest frame of the 
second scattering electrons. Multiple scattering of synchrotron photons
can be ignored. We obtain
$x_f=(\eta \epsilon_e/\epsilon_{B,f})$ 
for $(\eta\epsilon_e/\epsilon_{B,f})\ll 1$ and
$x_f=(\eta \epsilon_e/\epsilon_{B,f})^{1/2}$ for 
$(\eta\epsilon_e/\epsilon_{B,f})\gg 1$ (Sari \& Esin 2001).
$\mathcal{R}_M=\Gamma_d^2/\Gamma_0$ is the mass ratio defined in section
2.

The 1st IC peak is much lower than that of the forward shock as
$F_{max}^{IC}/F_{max,f} \lesssim 2\times10^{-4} ~\mathcal{R}_B^{-1/2}
n \Gamma_{0,2} R_{d,17}$.
Since the typical frequencies $\nu_{m,f}$ 
and $\nu_m^{IC}$ are around the X-ray band in the early afterglow phase, 
the forward shock emission should peak at a low cooling frequency 
$\nu_{c,f} \ll \nu_X$ (this could be achieved with 
a moderate density ambient medium), otherwise the SSC emission is masked 
by the forward shock emission.   

\section{Light Curves of SSC emission}
After a reverse shock crosses the shell, the hydrodynamic quantities of
a fluid element in the shell evolves as the Lorentz factor $\Gamma
\propto R^{-7/2}$, the mass density $\rho \propto R^{-3}\Gamma$ and  
the internal energy $e \propto \rho^{4/3}$ for a thick shell case, while
for a thin shell case they can be approximated as  $\Gamma \propto R^{-2}$, 
$\rho \propto (R^{-3}\Gamma)^{6/7}$ and  $e\propto \rho^{4/3}$
(Kobayashi \& Sari 2000). Using these scalings, one obtains that
the flux below the peak frequency $\min[\nu_m^{IC}, \nu_c^{IC}]$ 
evolves as $\sim t^{-1/2}$ for a thick shell and $\sim t^{-2/3}$
for a thin shell. The flux between $\nu_m^{IC}$ and $\nu_c^{IC}$ falls
as $\sim t^{-(5p+1)/5}$ for a thick shell and $\sim t^{-(3p+1)/3}$ for a 
thin shell. When $\nu_c^{IC} \sim t^{-2}$ becomes lower than the
observed frequency, the flux drops exponentially with time. However, the
angular time delay effect prevents an abrupt disappearance. The flux will
be determined by off-axis emission if the line-of-sight emission decays
faster than $t^{-2+\beta}$ where $\beta=-(p-1)/2$ or $-p/2$ is the
spectral index (Kumar \& Panaitescu 2000). 

The temporal index of a SSC flare in the rising phase is rather
uncertain, because it depends on the density profile of the fireball
shell. We assume a simple homogeneous shell. The hydrodynamic quantities
and the number of shocked electrons during shock crossing are given by 
$\Gamma \propto R^{-1/2}$, $e \propto \Gamma^2$, $\rho \propto
R^{-2}\Gamma^{-1}$ and $N_e \propto R^2$ for a thick shell,  
and $\Gamma\propto R^{-g}$, $e \propto \Gamma^2$ and 
$\rho \propto R^{-3}\Gamma^{-1}$ and $N_e \propto R^{3/2}$ for a thin
shell  where $g$ is a parameter (e.g. Kobayashi 2000). Around the
shock crossing, the shock becomes mildly relativistic (Sari \& Piran
1995; Kobayashi, Piran \& Sari 1999). We consider the behavior of 
a light curve near the peak $(g=1/2)$. The break frequencies of the
SSC emission evolve as $\nu_c^{IC}\propto t^{-3/2}$, 
$\nu_m^{IC} \propto t^{1/2}$ (thick shell) or $t^{5/2}$ (thin shell).
Since during the shock crossing $\nu_m^{IC}$ increases while $\nu_c^{IC}$ 
decreases, an interesting regime is $\nu_m^{IC} < \nu_{obs} <
\nu_c^{IC}$. Assuming $p=2.5$, for a thick shell the flux increases as
$t^{7/8}$ in this regime, and it is almost constant in the other
regimes. For a thin shell, the flux in this regime  increases as
$t^{15/8}$. If $\nu_m^{IC}$ (or $\nu_c^{IC}$) crosses $\nu_{obs}$, the
flux behaves as $t^{-5/6}$ (or $t^{9/8}$). If both break frequencies
pass the observed frequency $\nu_c^{IC}< \nu_{obs} < \nu_m^{IC}$, the
flux decays as $t^{-3/4}$.  

\section{X-ray Flare from Thin Shell}
\begin{figure}
\plotone{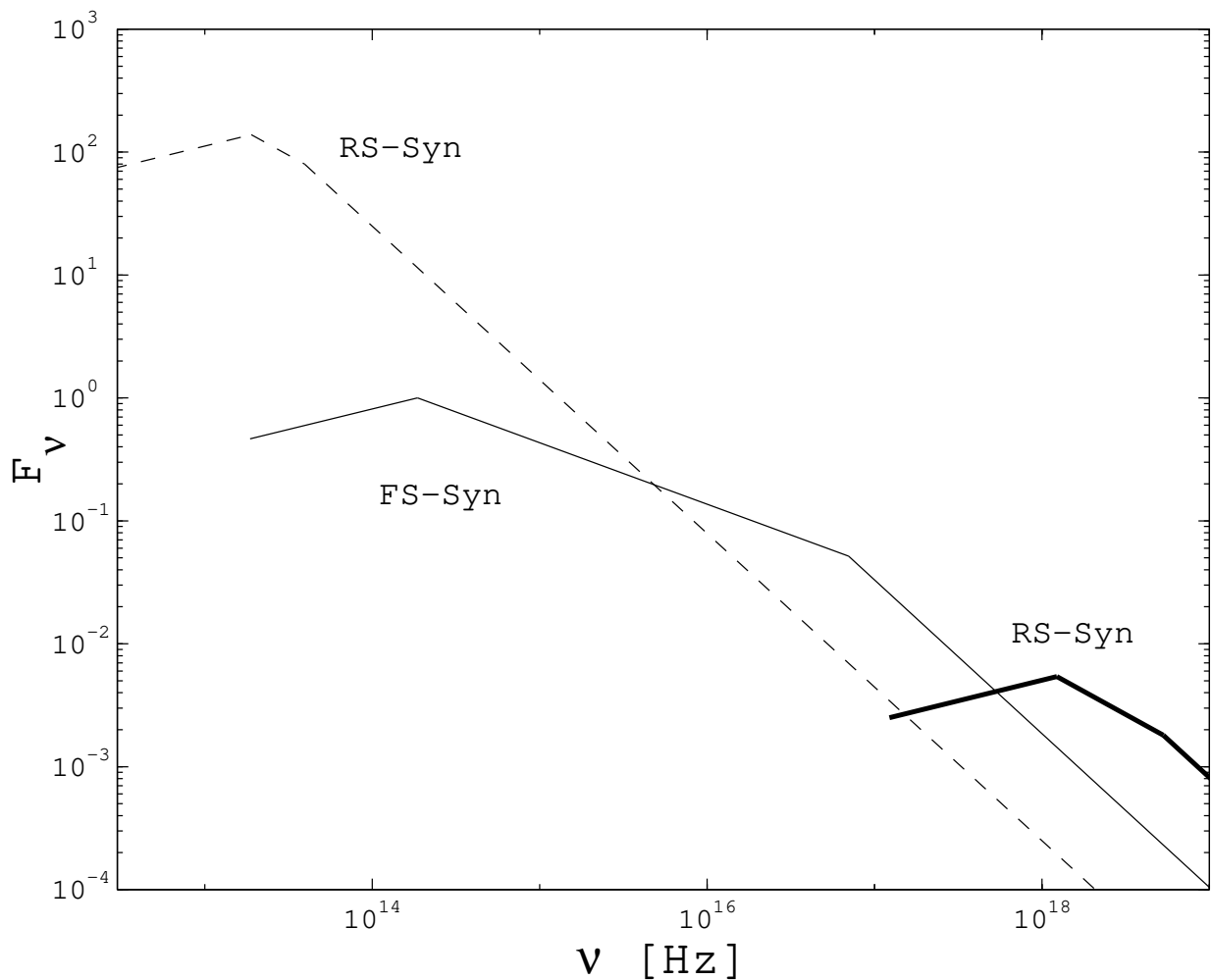}
\caption{Spectrum at the deceleration time:
Reverse Shock SSC (thick solid) and Synchrotron (thin dashed) 
emission, and Forward Shock Synchrotron emission (thin solid).
$z=1, E_{52}=1, \Gamma_0=80, n=5, \epsilon_e=0.3, \epsilon_B=0.03$
and $\epsilon_{B,f}=0.01$
The flux is normalized at the forward shock peak.
\label{fig:spec}}
 \end{figure}

\begin{figure}
\plotone{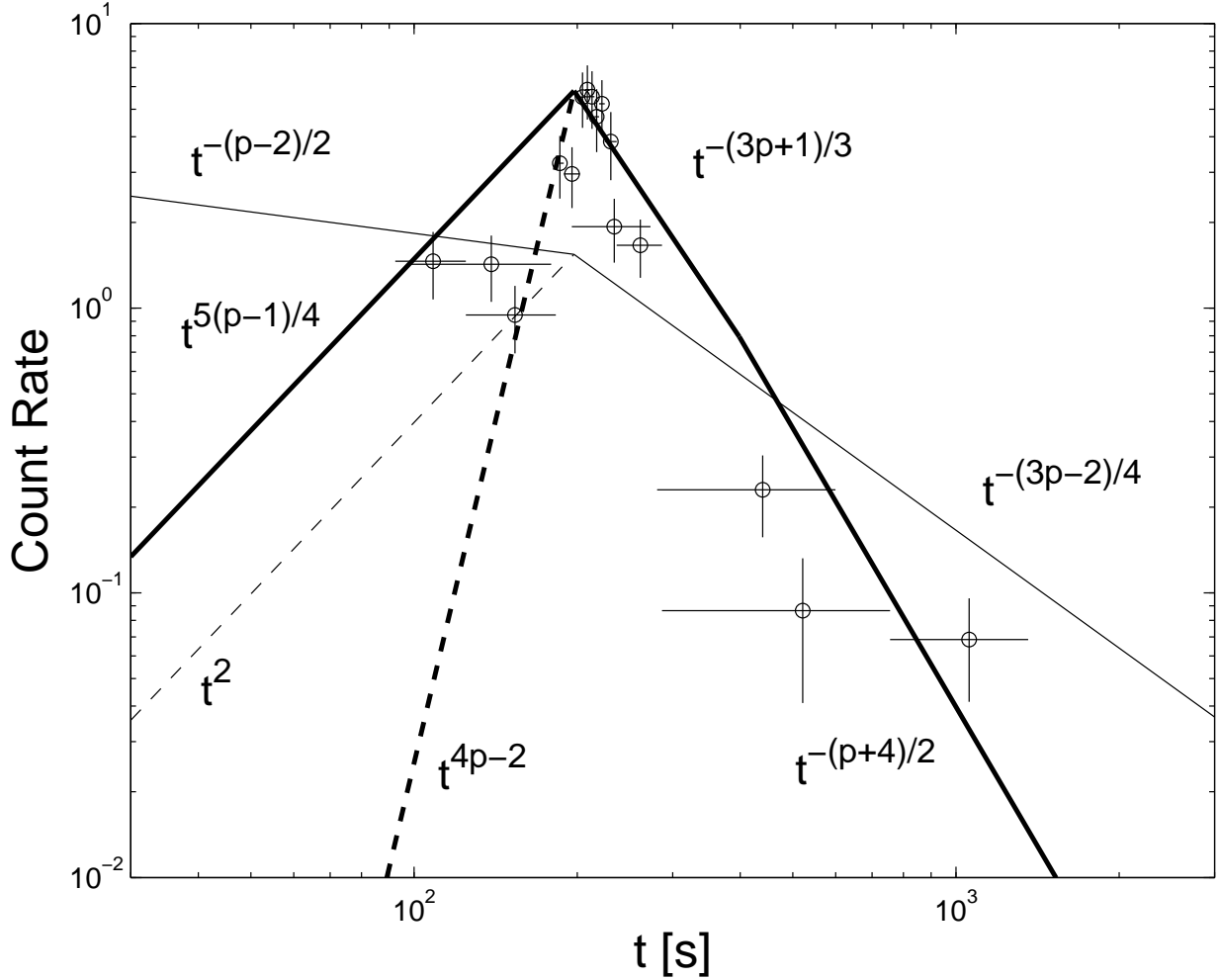}
\caption{X-ray Light Curve:
   Reverse Shock SSC (thick solid) and Forward Shock Synchrotron emission
  (thin solid).  The parameters are the same as in
  figure \ref{fig:spec} except $\Gamma_0=79$. The circles represent 
  the X-ray light curve (counts s$^{-1}$) of GRB 050406 from Romano et
  al. (2006). The theoretical light curves are normalized as the reverse 
  shock SSC emission at the deceleration time fits the
  observed peak (the peak counts $\sim 6$ counts s$^{-1}$). If the width 
  of the fireball shell is constant before the deceleration, the shock
  emissions rise more rapidly at the beginning (dashed lines). 
\label{fig:lc}}
 \end{figure}

In this section, we first study the typical thin shell case, and show the 
spectrum and light curve of the early afterglow. If the reverse 
shock is in the thin shell regime, the SSC peak is separated 
from the GRB emission. The deceleration time is 
$t_d  \sim 190 ~\zeta E_{52}^{1/3} (\Gamma_0/80)^{-8/3} (n/5)^{-1/3}$ sec.
The typical and cooling frequencies of the forward shock synchrotron
emission at the deceleration time are given (e.g. Sari, Piran \& Narayan
1998) by  
\begin{eqnarray}
\nu_{m,f} &\sim& 7.0 \times 10^{16} 
\zeta^{-1} 
\paren{\frac{\epsilon_{B,f}}{0.01}}^{1/2} 
\paren{\frac{\epsilon_{e}}{0.3}}^2
\paren{\frac{\Gamma}{80}}^4 
\paren{\frac{n}{5}}^{1/2}
~\mbox{Hz}, \\
\nu_{c,f} &\sim& 1.9 \times10^{14}  
\zeta^{-1} E_{52}^{-2/3} \paren{\frac{1+x_f}{6.5}}^{-2}
\paren{\frac{\epsilon_{B,f}}{0.01}}^{-3/2}
\paren{\frac{\Gamma}{80}}^{4/3} 
\paren{\frac{n}{5}}^{-5/6} 
~\mbox{Hz},
\end{eqnarray}
Using eq. (\ref{eq:zkm03}), the break frequencies of the
reverse shock synchrotron emission are 
\begin{eqnarray}
\nu_{m} &\sim& 1.9 \times 10^{13} 
\zeta^{-1} 
\paren{\frac{\epsilon_{B}}{0.03}}^{1/2} 
\paren{\frac{\epsilon_{e}}{0.3}}^2
\paren{\frac{\Gamma}{80}}^2 
\paren{\frac{n}{5}}^{1/2}
~\mbox{Hz}, \\
\nu_{c} &\sim& 3.9 \times10^{13}  
\zeta^{-1} E_{52}^{-2/3} \paren{\frac{1+x+x^2}{6.2}}^{-2}
\paren{\frac{\epsilon_{B}}{0.03}}^{-3/2}
\paren{\frac{\Gamma}{80}}^{4/3} 
\paren{\frac{n}{5}}^{-5/6} 
~\mbox{Hz},
\end{eqnarray}
where we assumed a larger value of $\epsilon_B$ for the fireball than
that of the blast wave. Using eq. (\ref{eq:IC}),
the break frequencies of the reverse shock SSC emission are
\begin{eqnarray}
\nu_{m}^{IC} &\sim& 1.2 \times 10^{18} 
\zeta^{-1} 
\paren{\frac{\epsilon_{B}}{0.03}}^{1/2} 
\paren{\frac{\epsilon_{e}}{0.3}}^4
\paren{\frac{\Gamma}{80}}^2 
\paren{\frac{n}{5}}^{1/2}
~\mbox{Hz}, \\
\nu_{c}^{IC} &\sim& 5.3\times10^{18}
\zeta^{-1} E_{52}^{-4/3}
\paren{\frac{1+x+x^2}{6.2}}^{-4}
\paren{\frac{\epsilon_{B}}{0.03}}^{-7/2}
\paren{\frac{\Gamma}{80}}^{2/3} 
\paren{\frac{n}{5}}^{-13/6} 
~\mbox{Hz}
\end{eqnarray}
We plot the broad band spectrum in figure \ref{fig:spec}. The reverse 
shock SSC emission (thick solid line) dominates X-ray band ($5$keV 
$\sim 10^{18}$ Hz) in this example case. Figure \ref{fig:lc} shows an
X-ray light curve around the deceleration time. The reverse shock SSC
emission produces an X-ray flare around the deceleration time.
For reference, we plot the X-ray light curve of GRB 050406. Compared to 
the the theoretical SSC flare (thick solid line), the observed flare
rises more rapidly, especially around the peak. As we gave a caveat in
the previous section, the temporal index in the rising phase is rather
uncertain. This sharp rise might be due to inhomogeneity of a fireball
shell (we have assumed a homogeneous shell to evaluate the light
curve). The internal shock model requires a highly irregular outflow
from the GRB central engine. Since the hydrodynamic interaction inside
the flow smooths the velocity and pressure profiles, but not the density
profile, fireball ejecta might have an irregular density profile at the
deceleration time. Emission from the ejecta during a reverse shock
crossing could reflect the light curve of the prompt emission produced
by internal shocks (Nakar \& Piran 2004). The duration of GRB 050406 was
$T_{90}=5\pm1$ s in the 15-350keV band and the light curve peak had a
fast rise, exponential decay (FRED) profile (Krimm et al. 2005). Around
the peak, a reverse shock might hit a higher density part of the shell. 

Another possibility is that a reverse shock might stay in the
Newtonian regime during the whole evolution. The SSC emission from a
Newtonian reverse shock is expected to rise as rapidly as $\propto
t^{4p-2}=t^8$ for $p=2.5$ (thick dashed line). In such a case, the
forward shock emission also increases faster (thin dashed line). This could
happen if later ejecta from the central engine injects additional energy
to the inner tail of the shell after the internal shock phase, and it
further smooths the velocity profile of the shell. When the spreading of
a shell width is not significant, a reverse shock does not evolve to
mildly relativistic at the deceleration time (Sari \& Piran 1995). X-ray
flares with a moderate amplitude like this event can be produced by the
reverse shock SSC process, although the sharp structure $dt/t<1$ might be
difficult to be explained in external shock related  models (Kobayashi
\& Zhang 2006).  

\begin{figure}
\plotone{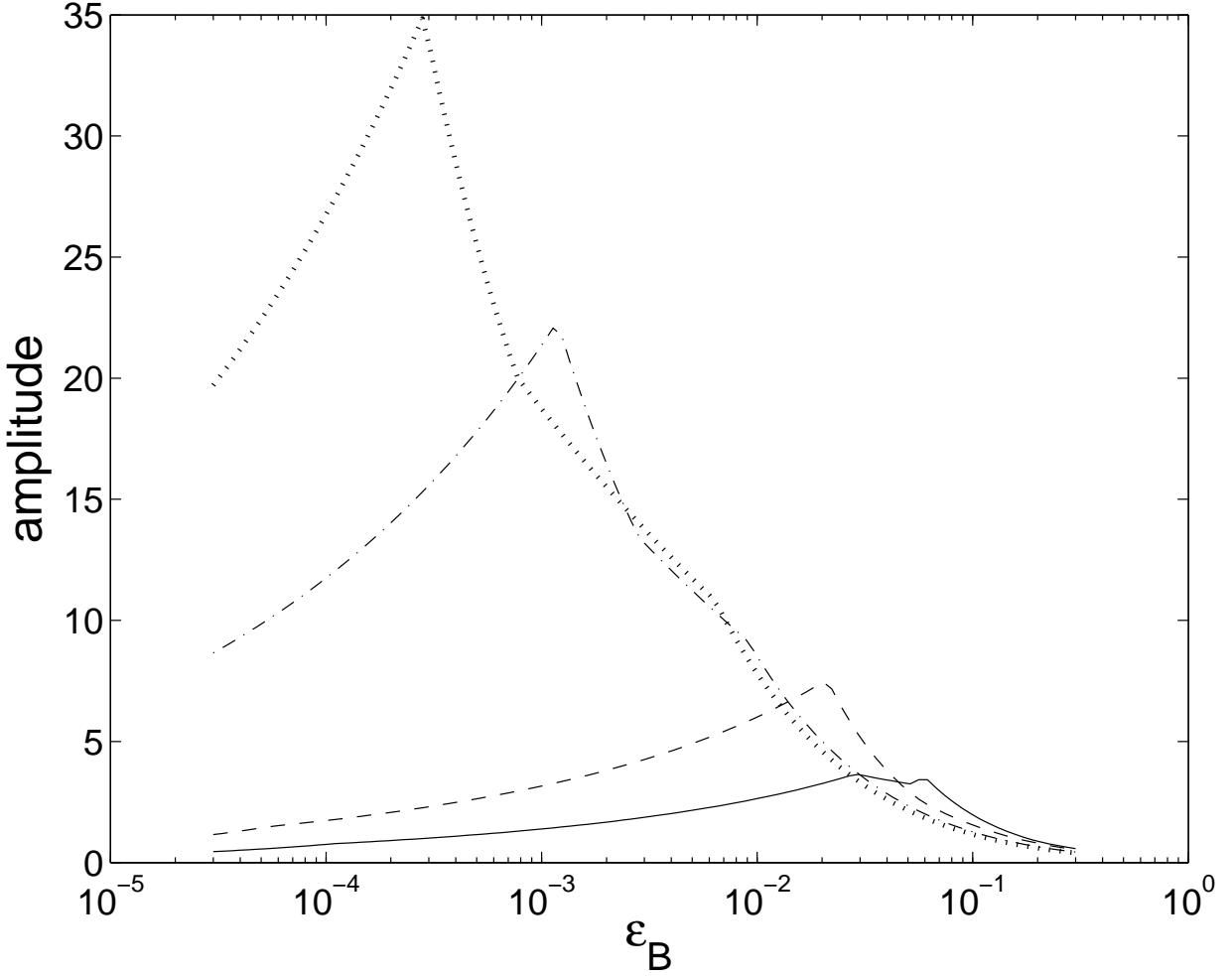}
\caption{X-ray flare amplitude:
Flux ratio between the reverse shock IC and 
forward shock synchrotron emission $F_\nu^{IC}/F_{\nu,f}$ at X-ray 
band (5keV) at the deceleration time is plotted as a function of
  $\epsilon_B$. $n=5$ (solid), 10 (dashed), 50 (dashed dotted) or 100
  (dotted).  $z=1, E_{52}=1, \epsilon_e=0.3$ and
$\epsilon_B/\epsilon_{B,f}=3$. For a smaller $\epsilon_B$, the Compton
  parameter $x$ is larger and the typical frequency of the IC emission
  $\nu_m^{IC}$ is lower. The amplitude of an X-ray flare peaks at a
  moderate $\epsilon_B$ with which $\nu_m^{IC}$ is close to the X-ray band.
\label{fig:amp}}
 \end{figure}

We consider how the relative amplitude of X-ray flares depends on 
parameters. Since X-ray flares occur a few hundred seconds
after the prompt emission, we will consider the thin shell case. Fixing
the deceleration time $t_d  \sim 190 ~\zeta E_{52}^{1/3} (\Gamma_0/80)^{-8/3}
(n/5)^{-1/3}$ sec, the initial Lorentz factor is a function of the ISM
density (we assume the typical explosion energy and redshift). 
A larger ratio of $\epsilon_e/\epsilon_B$ enhances the scattering 
process. Assuming the equipartition value $\epsilon_e=0.3$, we evaluate 
the relative amplitude of an X-ray flare as a function of $\epsilon_B$.
The results are shown for different values of $n$ in figure
\ref{fig:amp}. The amplitude is as large as a few tens if $\epsilon_B \ll 1$
and $n \gg 1$. However, a high density requires a low Lorentz factor
(i.e. $\Gamma\sim 55$ for $n=100$).

\section{Conclusions}
We have investigated the synchrotron self inverse-Compton (SSC) emission  
from the reverse shock. The synchrotron process is expected to produce
optical/IR photons, which are up-scattered into the X-ray band
by electrons heated by the reverse shock. For a thick shell, the X-ray
flare occurs at the end of the prompt gamma-ray phase. The emission
decays as $\sim t^{-2}$ or slightly steeper. For a thin shell, the emission
initially increases as $\sim t^2$ (the scaling could be significantly
different if a fireball shell is highly irregular), the peak should be
separated from the prompt emission, and after the peak the flux decays
as $\sim t^{-(3p+1)/3}$ or steeper. If off-axis radiation dominates, the 
temporal and spectral indices $L\propto t^{\alpha}\nu^{\beta}$ should
satisfy the relation $\alpha=-2+\beta$. 
A weakly magnetized fireball ($\epsilon_{B,f} < \epsilon_B \ll 1$) in a high
density ambient medium provides favorable conditions for producing a
significant X-ray flare. The contrast between the IC flare and the
baseline X-ray emission is at most one order of magnitude if the
synchrotron process dominates the electron cooling. If the IC dominates,
the contrast could be larger. However, since most of the energy is
radiated in the 2nd order IC component around 10-100 MeV, the X-ray bump
and the baseline X-ray emission are less energetic than the latter. The
optical flash  (due to synchrotron) is highly suppressed.

Recently Swift XRT detected X-ray flares in the early afterglows of
GRB 050406 and GRB 050502b (Burrows et al 2005). The afterglow of GRB
050406 brightens by a factor of 6 between 100 and 200 s post-burst
before starting on the rapid decline seen in other prompt X-ray 
afterglows. GRB 050502b had an even stronger X-ray flare, brightening by 
a factor of $\sim 10^3$ to a peak 700 s after the burst. Both afterglows 
were very faint at 100 s post-burst (a factor of a few - 100 fainter than 
previous XRT-detected afterglows). 

In the case of GRB 050502b, the X-ray flare contrast factor $\sim 10^3$ 
requires a Compton parameter $x$ larger than $\sim 100$, which means 
$\epsilon_e/\epsilon_B \gtrsim 10^8$. With plausible values of the other 
parameters, the SSC emission should appear at an energy band well above
X-rays. The very sharp profile $dt/t \ll 1$ also rules out external
shock models.  Another mechanism (e.g late time internal shocks; 
Burrows et al. 2005; Falcone et al. 2006; Romano et al 2006; 
Zhang et al. 2006; Nousek et al. 2006; Fan \& Wei 2005; Wu et al. 2005; Ioka, 
Kobayashi \& Zhang 2005) 
is likely to play a role in the production of the X-ray flare. 
On the other hand, the X-ray flare of GRB 050406 might be explained 
with the reverse shock SSC emission discussed here, although the
rather narrow feature $dt/t<1$ might disfavor the external shock
interpretation. If we apply the SSC model to the flare, the peak time
$t_d \sim 200$ sec gives an initial Lorentz factor $\Gamma_0 \sim 80
~(t_d/200\mbox{s})^{3/8}\zeta^{3/8}E_{52}^{1/8} (n/5)^{-1/8}$. 
The reverse shock SSC emission can not explain multiple X-ray
flares in a single event. Such behavior is observed in recent Swift
bursts (Burrows et al. 2005; Falcone et al. 2006; Romano et al. 2006; 
O'Brien et al. 2006).

In thick shell cases, the SSC flare rising portion overlaps in time 
with the prompt emission seen by BAT. Although it could be difficult 
to separate the X-ray emission of the prompt and the reverse SSC flare
rising components, the rapidly decaying portion of the reverse SSC 
flare could be detectable. Swift has reported many of the X-ray
afterglows detected have an early steep decay phase, before going into a
more common shallow decay. In some of these cases, the early steep decay
may be interpreted as the tail portion of a reverse shock SSC X-ray
flare. 

We note also that the reverse shock SSC mechanism predicts,
besides an X-ray flare, a strong GeV flare from 2nd order IC, and 
in some cases from first order IC. Thus, bursts with strong early
X-ray flares should be good candidates for GLAST.

We thank Patrizia Romano and the Swift team for providing XRT data. 
This work is supported by Eberly Research
Funds of Penn State and by the Center for Gravitational Wave Physics
funded by NSF under cooperative agreement PHY 01-14375 (for SK), 
NASA NNG04GD51G (for BZ),  
NASA AST 0098416 and NASA NAG5-13286 (for PM), 
and NASA Swift GI program (for BZ, SK and PM).
\clearpage

\noindent {\bf References}

Akerlof,C.W. et al. 1999, Nature, 398, 400. 

Beloborodov,A.M. 2005, ApJ, 618, L13. 

Burrows,D. et al. 2005, Science, 309, 1833.

Falcone,A.D. et al. 2006, ApJ, 641, 1010.

Fan,Y. \& Wei,D.M. 2005, MNRAS, 364, L42.

Granot,J.\&Guetta,D. 2003, ApJ, 598, L11. 

Ioka,K.,Kobayashi,S.\& Zhang,B. 2005, ApJ, 631, 429.

Kobayashi,S 2000, ApJ, 545, 807. 

Kobayashi,S, Piran,T. \& Sari,R. 1999, ApJ, 513, 669. 

Kobayashi,S \& Sari,R. 2000, ApJ, 542, 819. 

Kobayashi,S. \& Zhang,B. 2006, submitted to ApJ, astro-ph/0608132.

Kumar,P. \& Panaitescu,A. 2000, ApJ, 541, L51. 

Krimm,H. et al. 2005, GCN 3183

\Mesz,P. \& Rees,M.J. 1993, ApJ, 418, L59. 

M\'esz\'aros,P. \& Rees,M.J. 1997, ApJ, 476, 231. 

Nakar,E. \& Piran,T. 2004, MNRAS, 353, 647

Nousek,J.A. et al. 2006, ApJ, 642, 389.

O'Brien, P.T. et al. 2006, ApJ, 647, 1213.

Romano,P. et al. 2006 A \& A, 450, 59.

Roming,.P et al. 2006 ApJ in press, astro-ph/0509273. 

Sari,R. \& Esin,A.A. 2001 ApJ, 548, 787. 

Sari,R. \& Piran,T. 1995 ApJ, 455, L143. 

Sari,R. \& Piran,T. 1999 ApJ, 520, 641. 

Sari,R., Piran,T. \& Narayan,R. 1998 ApJ, 497, L17. 

Tagliaferri,G. et al. 2005 Nature, 436 985.

Wang,X.Y.,Dai,Z.G.\&Lu,T. 2001a, ApJ, 546,L33. 

Wang,X.Y.,Dai,Z.G.\&Lu,T. 2001b, ApJ, 556, 1010. 

Wu,X.F. et al. 2005, submitted to ApJ, astro-ph/0512555.

Zhang,B., Kobayashi,S. \& M\'esz\'aros,P. 2003, ApJ, 595, 950. 

Zhang,B. \& Kobayashi,S.  2005, ApJ, 628, 315. 

Zhang,B. et al. 2006, ApJ, 642, 354.
\clearpage

\clearpage

\clearpage

\end{document}